\documentclass[12pt,letterpaper]{iopart}

\usepackage{graphicx}       
\usepackage{amsfonts}       
\usepackage{verbatim}       
\usepackage{hyperref}       

\begin{document}


\newcommand{\shorttitle}{Demonstration of microscale optics in surface-electrode ion traps}
\title[\shorttitle]{Demonstration of integrated microscale optics in
   surface-electrode ion traps}
\author{J True Merrill$^1$, Curtis Volin$^2$\footnote{Author to whom any correspondence should be addressed.}, David Landgren$^2$, Jason M Amini$^2$, Kenneth Wright$^2$, S Charles Doret$^2$, C-S Pai$^2$, Harley Hayden$^2$, Tyler Killian$^2$, Daniel Faircloth$^2$\footnote{Current address: Ierus Technologies, 9122 Loxford Street, Lithia Springs, GA 30122-6413, USA.}, Kenneth R Brown$^1$, Alexa W Harter$^2$ and Richart E Slusher$^2$}

\address{$^1$ Schools of Chemistry and Biochemistry, Computational Science and Engineering, and Physics, Georgia Institute of Technology, Atlanta, GA 30332,
USA.}
\address{$^2$ Georgia Tech Research Institute, Atlanta, GA 30332, USA.}
\ead{curtis.volin@gtri.gatech.edu}

\begin{abstract}
In ion trap quantum information processing, efficient fluorescence collection is critical for fast, high-fidelity qubit detection and ion-photon entanglement.  The expected size of future many-ion processors require scalable light collection systems.  We report on the development and testing of a  microfabricated surface-electrode ion trap with an integrated high numerical aperture (NA) micromirror for fluorescence collection.  When coupled to a low NA lens, the optical system is inherently scalable to large arrays of mirrors in a single device.  We demonstrate stable trapping and transport of $^{40}$Ca$^+$ ions over a 0.63 NA micromirror and observe a factor of 1.9 enhancement in photon collection compared to the planar region of the trap.
\end{abstract}

\pacs{37.10.Ty, 03.67.Lx, 42.15.Eq}

\tableofcontents
\renewcommand{\leftmark}{\shorttitle}

\section{Introduction}

Arrays of trapped atomic ions are a promising system for implementing quantum information processing and quantum simulation.  The essential components of a universal ion-based quantum computer have been realized \cite{Blatt-Nature.453.1008.2008}, and the current challenge is to obtain quantum control over ionic qubits in a scalable manner. A practical method for increasing system complexity is the use of microfabricated surface-electrode traps \cite{Amini-NJP.12.033031.2010,Kim-QIC.5.515.2005,Chiaverini-QIC.96.253003.2005}, where lithographic manufacturing processes can produce complex electrode geometries \cite{Amini-NJP.12.033031.2010,britton-apl.95.173102.2009,Seidelin-PhysRevLett.96.253003.2006,Labaziewicz-PhysRevLett.100.013001.2008,Brown-PhysRevA.75.015401.2007,Leibrandt-QIC.9.0910.2009,Splatt-NJP.11.103008.2009}.  Surface-electrode trap architectures are well suited for integrating structures that perform specialized tasks such as qubit operations, ion transport, and memory \cite{Kielpinski-Nature.417.709.2002}.  In particular, optical elements can be integrated into the trap to improve state detection. 

Ion qubit state-detection relies on efficient collection of laser-induced ion fluorescence \cite{Burrell-PhysRevA.81.040302.2010, Myerson-PhysRevLett.100.200502.2008}.  Measurement times for high-fidelity readout are set by the collection efficiency, frequently limited by a small light collection solid-angle.  An advantage of fast qubit measurement times is more effective control of quantum states \cite{Barreiro-Nature.470.486.2011} with applications to quantum error correction protocols \cite{Chiaverini-Nature.432.602.2004} and measurement-based quantum computers \cite{Raussendorf-NJP.9.199.2007}. Increased light collection may also be useful for entangling distant ions by photon interference \cite{Maunz-PhysRevLett.102.250502.2009}, provided sufficient spatial coherence exists between the collected photons for efficient single-mode fibre coupling.

As the system size is increased, it will be necessary to detect the states of multiple ions simultaneously in order to keep operation times low.  With current bulk-optics systems, optimizing the light collection from a single ion can restrict the field of view (FOV), thus limiting the ability to perform parallel measurements.  Several methods for scalable state detection have been investigated, such as direct coupling to optical fibres \cite{VanDevender-PhysRevLett.105.023001.2010}, microfabricated Fresnel optics \cite{Streed-PhysRevLett.106.010502.2011, Brady-arxiv1008.2977}, and optical cavities.  Recently, surface-electrode traps have been fabricated directly on a planar mirror, resulting in two ion images \cite{Herskind-arxiv.1011.5259}.  The largest collection efficiency thus far has been demonstrated with a macroscopic three-dimensional spherical mirror trap \cite{Shu-PhysRevA.81.042321.2010}.  

Here we examine a multi-scale fluorescence collection system where high numerical aperture (NA) micromirrors are coupled to a macroscopic, low NA lens for efficient light collection over a large FOV.  An array of these mirrors could be integrated into a large trap, permitting simultaneous collection of light from many ions.  We designed and characterized an Al spherical mirror microfabricated into an Al-on-SiO$_2$ surface-electrode trap. The process is compatible with the standard fabrication technique described in detail in Section \ref{section:process} and similar to \cite{Leibrandt-QIC.9.0910.2009}.  

This paper is organized as follows: section \ref{section:design} describes the design of the microfabricated mirror trap and optics, section \ref{section:process} describes the trap architecture and fabrication procedure, and section \ref{section:data} presents measurements of the collection enhancement of a trapped atomic ion over the mirror.  Section \ref{section:conclusion} concludes with proposed improvements and potential applications.

\section{Trap and optics design}
\label{section:design}

\begin{figure}
\begin{center}
\includegraphics{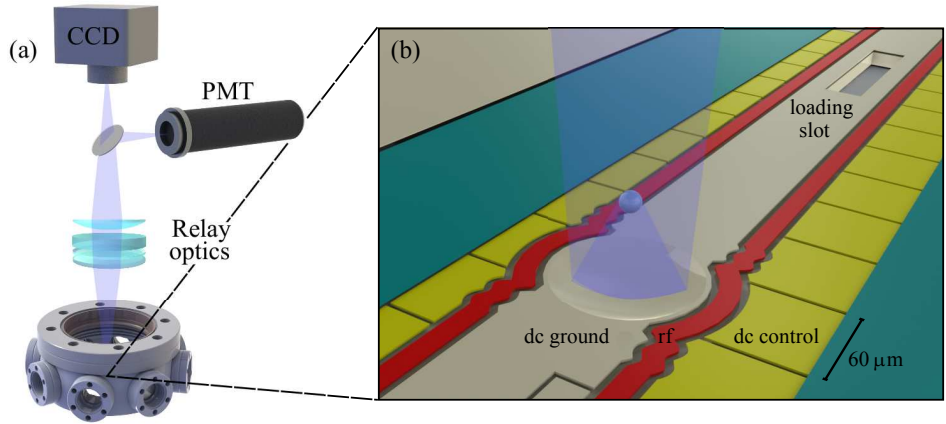}
\end{center}
\caption{(a) Diagram of the experimental apparatus.  The ion trap is mounted on a CPGA carrier placed in an ultra-high vacuum chamber with laser access across the surface of the device.  Scattered fluorescent photons from a trapped $^{40}$Ca$^+$ ion collected by a relay optic are detected by a CCD camera and a PMT. (b) Trap layout showing the integrated miromirror, rf rails, and the dc control electrodes.  The micromirror improves collection efficiency by locally increasing the collection solid-angle.}
\label{experiment}
\end{figure}

The design consists of a surface-electrode Paul trap with an integrated reflective mirror component for improved photon collection.  The surface trap is mounted on a CPGA carrier and placed in an ultra-high vacuum chamber (figure \ref{experiment}a).  A radio-frequency (rf) potential applied to a pair of rail electrodes confines the ions radially  (figure \ref{experiment}b).  Axial confinement is achieved by biasing a subset of the 42 independent dc control electrodes which lie along the rf rails adjacent to the trapping region.  Typical trapping potentials utilize five electrodes per side to generate harmonic wells with secular frequencies between 0.5 and 2 MHz.  By slowly varying the potentials applied to the control electrodes, a trapped ion can be smoothly transported to distinct regions of the device, including to regions that contain specialized structures for efficient state readout.  

An approximately spherical micromirror directly integrated into the central dc electrode reflects a large NA cone of fluorescence from an ion near the focus. Any mirror misalignment or deviation from an ideal profile results in a divergent cone of light.  The alignment of the optical focus and the ion is entirely the product of design and microfabrication of the trap and is not sensitive to thermo-mechanical misalignment.  A macroscopic relay optic is placed outside the vacuum chamber to collect and focus fluorescent light onto a detector. In this role, rather than imaging the ion, the relay lens images the micromirror onto the detector.  As such, the relay lens spot size is only required to match the selected detector size, making it simple to design and assemble.  The relay optics are designed to be tolerant of misshaped and misaligned micromirrors.  Furthermore, this multi-scale light collection system can direct light from a multitude of micromirrors distributed across a large FOV to independent detectors.

\subsection{Designing traps for micromirror integration} 

For compatibility with VLSI fabrication techniques \cite{Leibrandt-QIC.9.0910.2009}, we consider only designs in which no electrode edges are patterned inside the micromirror and where the electrode forming the mirror surface is grounded.  We estimate the solid-angle coverage provided by the mirror and its dependence on the electrode configuration by examining two analytical models in the gapless plane electrostatic approximation \cite{Wesenberg-PRA.78.063410.2008}: a ring trap geometry which enables a high collection efficiency of reflected photons and a linear-strip electrode geometry compatible with ion shuttling.  These models represent the design extremes; features of both designs are combined in a hybridized wrapped-electrode geometry which is numerically optimized to minimize the influence of the mirror on the pseudopotential tube.

First, consider a trap geometry where the mirror is surrounded by a narrow rf ring electrode.  Neglecting the depression of the mirror cavity (mirror sag), we use the gapless plane electrostatic approximation to calculate the inner radius of the rf rail $r$ (equal to half the mirror diameter) as a function of the ion height $h$ above the trap surface
\begin{equation}
r = h\left[ \frac{3}{4} \sin^{-2}\left(\frac{\pi}{6}+\theta\right)-1 \right]^{1/2}, \: 0 < \theta < \pi/6,
\end{equation}
where $\theta$ is proportional to the angular width of the rf electrode as seen from the ion (see \cite{Wesenberg-PRA.78.063410.2008}).  We immediately note that the collection angle $\varphi = \arctan(r/h)$ does not depend on the mirror shape or size. The upper bound ($\theta = 0^{\circ}$, corresponding to an rf electrode with an infinitesimal width) gives $\varphi=54.7^{\circ}$ (NA = 0.82, 21$\%$ geometric collection efficiency).  For a reasonable but small rail angle, $\theta=16^{\circ}$, $\varphi=50^{\circ}$ (NA = 0.76, 18$\%$ geometric collection efficiency).  These analytic results neglect the effect of the micromirror depression on the fields. The influence of the micromirror on the trapping fields will lower the ion height and is therefore expected to somewhat improve the collection efficiency.  Moving any portion of the rf rails away from the mirror will raise the ion height, therefore the insertion of small isolation gaps between electrodes as required for real traps will reduce collection efficiency. 

\begin{figure}
\begin{center}
\includegraphics{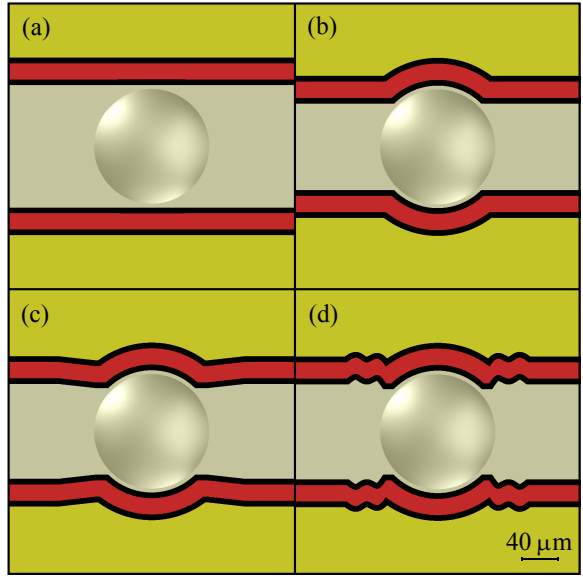}
\end{center}
\caption{Illustration of the geometry of the trap and micromirror for various design iterations. (a) a trap with linear rf rails tangent to the mirror, (b) rf rails wrapped around the mirror, (c) rf rails with a pinch and taper to a linear section, (d) the final geometry with genetic algorithm optimized rf electrodes.}
\label{mirror_design}
\end{figure}

While the maximum collection efficiency is achieved with a surface ring trap, we require a mirror compatible with a scalable architecture that allows for the shuttling of ions. Figure \ref{mirror_design}a shows an example of such an architecture in which the mirror is tangent to the rf rails of a linear section. In this design, any ion in the linear section may be transported over the mirror for readout.  Once again using the gapless plane electrostatic approximation, we find the relationship between the ion height and mirror radius for this configuration,
\begin{equation}
r=h\tan(\pi/4-\theta).
\end{equation}
Again, $r/h$ is a function only of the rf electrode angle, and the radius of curvature (ROC) of the mirror is linearly proportional to the ion height. The upper bound for the mirror acceptance angle, $\varphi=45^{\circ}$, is again found when $\theta=0^{\circ}$ (NA = 0.71, 15$\%$ collection efficiency). For a reasonable rail angle, $\theta=4^{\circ}$, $\varphi=41^{\circ}$ (NA = 0.66, 12$\%$ geometric collection efficiency). 

Figure \ref{mirror_design}b shows a concept for a hybrid of the two analytic models that improves the collection efficiency by wrapping the rf rail around the mirror. We choose a conservative design, with a target ion height of 63 $\mu$m, $\theta = 16^{\circ}$, and a $45^{\circ}$ wrapping of the rf rail around the mirror. Approximate values of the mirror ROC (150 $\mu$m) and radius (60 $\mu$m) were found from the above analytic forms after including a 6 $\mu$m gap around the edge of the rf electrodes and a 4 $\mu$m flat shelf around the edge of the mirror. The resulting optimal micromirror sag is 12 $\mu$m and the NA is 0.69 (geometric collection efficiency 14$\%$).  These parameters represent an idealized target geometry for fabricated micromirrors.  Errors in microtrap fabrication may lead to significant deviations in the mirror profile.

We performed numerical simulations on these candidate designs using our own method-of-moments electrostatics code.  Similar methods were used to optimize trapping electrode geometries.  Starting with the design in figure \ref{mirror_design}b, the rf rail width was adjusted until the mirror focus was aligned with the pseudopotential null (17 $\mu$m rail width) while keeping the inner edge of the rail and mirror profile constant.  Next, the spacing between the rf rails in the linear section was optimized so that the ion height over the linear section approximately matched the height in the mirror.  A pinch in the rf electrodes was inserted at the transition region between the linear and wrapped rail geometries (figure \ref{mirror_design}c) to reduce variation in the ion height.  This adjustment did not change the location of the rf null over the mirror. 

\begin{figure}
\begin{center}
\includegraphics{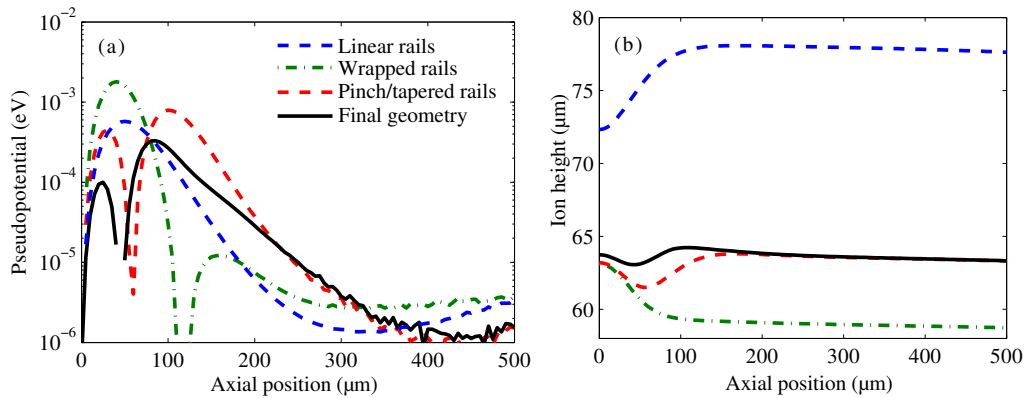}
\end{center}
\caption{(a) Residual pseudopotential at the minimum in the transverse plane as a function of axial displacement from the mirror centre for various design iterations.  The residual pseudopotential vanishes at points where the confining fields cancel.  (b) Height of the calculated pseudopotential minimum as a function of position along the trap axis.  Far from the mirror, the pseudopotential height asymptotically approaches a limiting value controlled by the rf rail spacing in the linear region.}
\label{pseudopotential}
\end{figure}

Finally, a genetic algorithm was used to optimize the rf rail geometry near the wrapping region. The algorithm uses a fitness function that minimizes the rf-noise motional heating rate \cite{Blakestad-PRL.102.153002.2009} while maintaining a nearly uniform ion height down the linear section.  The fitness function is proportional to $\int_C [\partial E^2(z) / \partial z ]^2 dz$, where $E(z)$ is the applied electric field and the contour $C$ follows the pseudopontential minimum along the axial coordinate $z$.  Perturbations to the rail geometry are parametrized by a set of edge points which are systematically displaced from the input geometry.  A spline function was used to interpolate the rf rail edge between these points.  The optimizer was allowed to adjust the number of points and the distance of the points from the axial centre line of the trap, under the constraint that the width of the rf rail was held constant and the rf null location was maintained within 0.25 $\mu$m of the mirror focus.  The resulting optimization produced several candidate solutions. The solution with the fastest decline in the rf field along the trap axis was selected for fabrication. This design is shown in figure \ref{mirror_design}d.

\subsection{Design of relay optics for scalable state detection}
\label{section:relay}

\begin{figure}
\begin{center}
\includegraphics{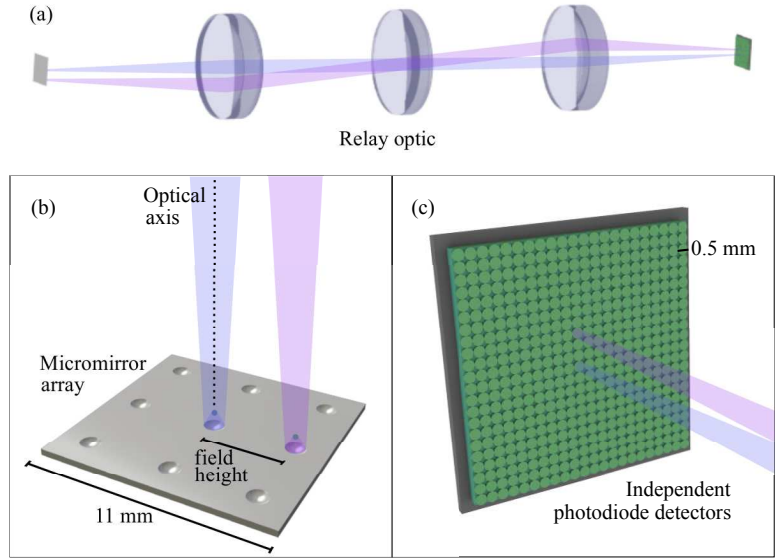}
\end{center}
\caption{(a) Diagram of a multi-scale light collection system for a proposed trap with an array of mirrors.  Collected fluorescence from each individual micromirror (b) is relayed to an independent detector (c).  The relay lens assembly can image mirrors over the entire 11 $\times$ 11 mm$^2$ trap chip with minimal cross-talk between detectors.  For clarity, the trap electrodes have not been drawn and only a small number of mirrors were included.  The micromirrors are not drawn to scale.}
\label{multiscale}
\end{figure}

An important component of the multi-scale detection system is the macroscopic relay lens assembly.  In a proposed trap design with an array of mirrors, the relay lens directs light reflected from each micromirror site located to an independent sensor (e.g., a single element in a PMT or APD array) in the image plane, allowing independent, asynchronous readout for each mirror (see figure \ref{multiscale}).  The required complexity of the relay optic depends on the mirror spacing in the ion trap and the required insensitivity of the system to misalignment.

To illustrate the simplicity and robustness of this approach, a 1:1, NA = 0.14 relay lens was designed using 2'' diameter stock plano-convex lenses in a commercial optics simulation package (Zemax \textregistered).  A full technical description of the relay lens, including ray-tracing simulations, is provided in the supplemental materials.  In the simulations, a spherical mirror matching the ideal target geometry is placed at the object point of the relay optic.  The optic is designed so that light collected from each measurement region is focused onto independent 0.25 mm radius detectors in the image plane.  Individual micromirrors distributed anywhere on the $11 \times 11$ mm$^2$ chip are resolvable with zero cross-talk so long as no two mirrors are placed closer than $0.5$ mm centre-to-centre.  This distance corresponds to six dc control electrode widths in the current trap (see figure \ref{experiment}b).

\begin{figure}
\begin{center}
\includegraphics{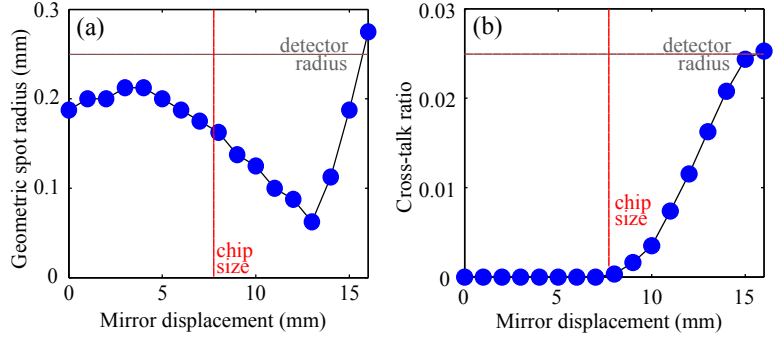}
\end{center}
\caption{(a) Simulated geometric spot radii of the ray bundle specularly reflected from a spherical micromirror for transverse displacements of the mirror from the relay optic axis (field height). Guides illustrating the detector radius and maximum field height set by the chip dimensions ($\sqrt{2} \times 11/2$ mm) are provided.  (b) Cross-talk ratio (fraction of collected light that reaches the image plane outside of the detector radius) versus mirror displacement from the relay optic axis.  The ratio is an upper bound on the cross-talk; in practice the actual cross talk will be smaller.  The calculation includes rays directly emitted by the ion and assumes 85\% mirror reflectivity with negligible transmission losses.}
\label{optics-tolerance1}
\end{figure}

\begin{figure}
\begin{center}
\includegraphics{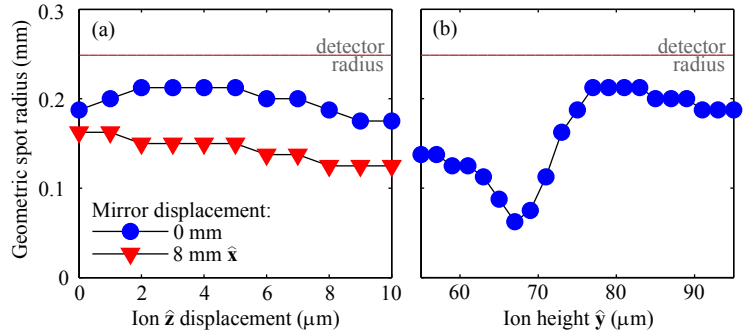}
\end{center}
\caption{(a) Simulated specularly reflected spot radii for ion displacements along the trap axis $\hat{z}$ from a mirror centre.  The reflected ray distribution is calculated for mirrors at field heights 0 mm and 8 mm.  The position of the detector is not adjusted to compensate for misalignment of the ion to the mirror.  (b) Spot radii for vertical ion displacements along the $\hat{y}$ axis from the vertex of a mirror located on the relay lens axis. }
\label{optics-tolerance2}
\end{figure}

To estimate the tolerance of the light collection system under various misalignments, we calculate the distribution of directly emitted and specularly reflected rays projected onto the detector plane.  We consider transverse displacements of the micromirror from the optical axis of the relay lens (field height).  In figure \ref{optics-tolerance1}a, we show that for mirror displacements exceeding the dimensions of the trap chip the light specularly reflected by the micromirror forms a localised spot less than 0.25 mm in radius in the image plane (the lens does not vignette any light reflected from the mirror).  The image plane is assumed to be distortion free.  That is, if the mirror is placed 8 mm from the axis of the relay lens, the detector is placed 8 mm from the axis in the opposite direction. 

Photons directly emitted by the ion are not well collimated and may arrive at one of several detectors, introducing a cross-talk error during state detection.  The relay optics have been designed to reduce this effect.  Figure \ref{optics-tolerance1}b shows that for all transverse mirror displacements within the chip dimensions, light from an individual ion relayed by the lens may be collected by a single 0.25 mm radius detector with a cross-talk of less than 0.05\%. Therefore, the designed multi-scale detection system is efficient for mirrors placed anywhere within the design FOV of the relay lens.  Note that larger, denser arrays of micromirrors are possible as fast, high fidelity readout is still possible with non-negligible cross-talk \cite{Myerson-PhysRevLett.100.200502.2008}.

We also consider misalignments of the ion relative to the micromirror either in the transverse $\hat{x}$-$\hat{z}$ plane (e.g., the ion is not centred on the optical axis) or longitudinally (e.g., the mirror ROC is incorrect, the mirror sag is incorrect, or the rf rails are incorrectly sized). The magnitudes of these misalignments are bounded by the characteristic errors from trap fabrication. We conservatively estimate the transverse misalignment to be less than 4 $\mu$m and the mirror sag error to be less than 3 $\mu$m. Nondestructive measurement of the micromirror profile outside NA = 0.3 is not currently possible, leading to uncertainty in the longitudinal alignment of the mirror focus with the rf null. However, it is possible to accurately measure the sag of the mirror and post-select a chip that is closest to the design objective (within 1 $\mu$m). In figures \ref{optics-tolerance2}a-b the specularly reflected spot radius is plotted as a function of transverse and longitudinal misalignment. We find that the design performance of the collection system is not degraded for misalignments within the expected fabrication tolerances.  Finally, we note that increased FOV, improved misalignment tolerances, and other detector sizes and configurations may be accessible by redesign of the relay lens.

\section{Trap architecture and fabrication}
\label{section:process}

\begin{figure}
\begin{center}
\includegraphics{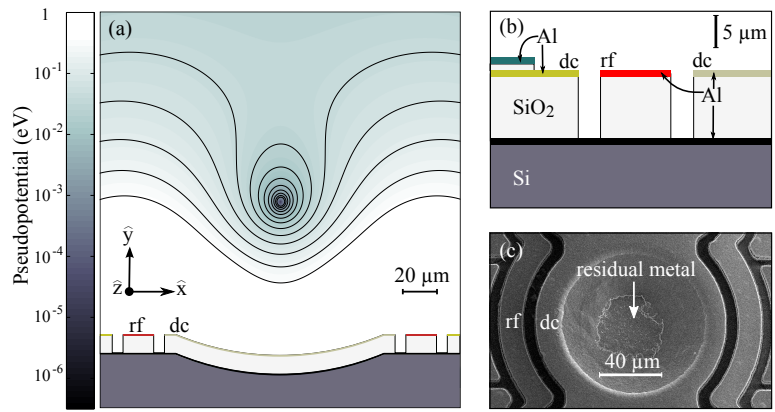}
\end{center}
\caption{(a) Trap cross section along the $\hat{x}$ radial direction at the centre of the mirror.  A logarithmic plot of the pseudopotential including equipotential lines is superimposed.  (b) Detail of trap fabrication on a silicon substrate.  For clarity, only features in the vertical direction are drawn to scale.  The trapping electrodes are isolated from the Si substrate by a 1 $\mu$m Al ground plane and 10 $\mu$m of insulating SiO$_2$.  Two additional patterned metal layers separated by 1 $\mu$m SiO$_2$ define the trapping electrodes (2nd level) and the integrated capacitive filters (3rd level).  (c) SEM image of the mirror on the prototype trap used in the experiment.  A lithography error in the final metal patterning step left 1 $\mu$m of residual aluminium from the capacitor layer in the centre of the mirror.}
\label{cross_section}
\end{figure}

The ion trap fabrication process is based on well established silicon VLSI processing techniques which enable the production of complex scalable structures.  The traps are fabricated on the surface of a $\langle$100$\rangle$ p-doped Si substrate and use sputtered Al electrodes with PECVD SiO$_2$ dielectric layers. Architecturally the device is similar to traps reported in \cite{Leibrandt-QIC.9.0910.2009} with several important design improvements, including integrated capacitive filters ($\sim$ 60 pF) to reduce rf pickup on the dc electrodes and asymmetric dc electrodes to simplify rotation of the secular axis for effective laser cooling. The design also features a through-chip loading slot.  

Fabrication begins by defining the mirror profile in the silicon substrate.  The process for producing recessed micromirrors in silicon uses an HF, HNO$_3$ and acetic acid (HNA) solution to etch isotropically through circular apertures patterned on a LPCVD silicon nitride mask \cite{Trupke-Appl.Phys.Lett.87.211106.2005,Albero-OpticsExpress.17.6283.2009,Noek-OpticsLett.14.2460.2010}. The wafer is etched in a room temperature HNA bath without agitation for 22 minutes after which the nitride mask is removed in HF.  The overall etch rate and final surface morphology are highly dependent on the concentrations of each of the etchant components and must be carefully optimized to provide a smooth, controllable etch \cite{Irving-J.Electrochem.Soc.107.1020.1960}. We have selected a 1 HF : 8 HNO${}_{3}$ : 1 CH${}_{3}$COOH (by vol.) solution for an etch with an anisotropy of $\sim 10\%$ \cite{Svetovoy-J.Micromech.Microeng.11.2344.2007} and a low occurrence of surface defects.  We find that the mirror diameter and radius of curvature can be controlled to within $\pm 2$ $\mu$m by choosing the appropriate circular aperture size and HNA etch time. Following wet processing, the frontside of the wafer is thinned using a combination of lapping and chemical mechanical polishing techniques to independently control the mirror sag.  The mirror surface is protected during thinning with a sacrificial 15 $\mu$m SiO$_2$ layer.

Following mirror fabrication, the trapping electrode structures are patterned over the polished substrate.  The spherical mirror profile is translated to the surface electrodes during build-up.  A cross-section of the device is shown in figure \ref{cross_section}b.  To prevent coupling of the trapping electrodes to the lossy Si substrate, a 1 $\mu$m Al ground plane and a thick 10 $\mu$m SiO$_2$ dielectric layer are deposited over the substrate.  Rf and dc electrodes are lithographically patterned and plasma etched from a 1 $\mu$m Al film deposited above the insulating oxide surface.  This layer of aluminium also serves as the mirror coating; the mirror surface itself is part of the central dc electrode.  A final pair of 1 $\mu$m thick SiO$_2$ and aluminium layers are patterned to form on-chip capacitive filters for grounding rf potentials on the control electrodes.  Isolation trenches separating the trapping electrodes are formed with a plasma etch which removes exposed SiO$_2$ between the electrode structures.  The removal of excess oxide from the trap surface also reduces sites where stray charges may accumulate and perturb the trapping potential.  An ICP Bosch process \cite{IntroToFab} is used to etch the loading slot through the substrate, resulting in a nearly vertical etch profile.

An SEM image of the micromirror in the prototype device used in this study is shown in figure \ref{cross_section}c.  During fabrication, a lithography error resulted in the incomplete removal of the final metal level in the centre of the mirror.  The presence of this rough ($\sigma_{RMS} > 40$ nm, measured by a Veeco Dimension 3100 AFM in tapping mode) residual metal is expected to severely degrade the optical performance of the device.  Among the batch of roughened mirrors, we selected traps for testing by the mirror surface finish rather than the geometry of the mirror profile.  For the trap tested in this study, the mirror geometry (ROC = 178 $\mu$m, $r = 50.5$ $\mu$m, NA = 0.63) differs substantially from the ideal profile.  Despite the poor quality of the mirror, the device was still able to demonstrate significant photon collection improvement (factor of 1.9 enhancement).  We expect to demonstrate larger collection enhancements as our fabrication processes are improved.

\section{Trapping and demonstration of collection enhancement}
\label{section:data}
The integrated mirror structure is characterized with single ions of $^{40}$Ca$^+$ fluorescing on the 397 nm $4^2$S$_{1/2} \leftrightarrow$ $4^2$P$_{1/2}$ cycling transition.  An additional optical repumping laser at 866 nm is used to prevent population trapping in the metastable $3^2$D$_{3/2}$ manifold \cite{Urabe-jjap.30.1532.1991,Haffner-review}.  Ions are loaded into the trap 600 $\mu$m from the mirror by photoionization of neutral $^{40}$Ca flux entering through the backside loading slot, preventing the deposition of metallic calcium over the trap surface.  Stray photoelectrons may charge exposed insulators inside the vacuum chamber, affecting trapping potentials.  A mesh ground plane 4 mm above the trap surface shields ions from stray fields while allowing the transmission ($T \geq 80$ \%) of fluorescent photons.  The rf trapping potential ($V_0 \approx 200$ V, $\Omega_{\mathrm{rf}} = 2 \pi \times 62.3$ MHz) is applied by a waveform generator filtered by a helical resonator.  Radial and axial mode frequencies were measured to be $2 \pi \times$(2.9, 2.2, 1.0) MHz. 

Ion shuttling is achieved by applying a set of slowly varying transport potentials ($|V| \leq 6 $V) to the dc control electrodes, producing a moving pseudopotential well which may be held stationary along any axial position in the trap.  The control potentials are applied by independent digital-to-analog converters with an update rate of 500 kHz. Typical shuttling operations include 10$^3$ transport potential update steps and last approximately 2 ms, with a success probability $P \geq 99.98 \%$.  After shuttling, a computer controlled piezo driven mirror steers the 397 nm cooling beam (aligned at a 45$^\circ$ angle from the trap axis) to track the ion.  The 866 nm beam is aligned to illuminate the entire trap axis.  

In the multi-scale approach, the collection enhancement from reflected light is controlled by the micromirror, while the FOV is dependent on the relay optics.  While an ideal relay optic for simultaneous readout over many distributed mirrors is described in section \ref{section:relay}, for demonstrating collection enhancement over a single mirror, we elected to use the relay optic system already present on the apparatus (1:10, design NA = 0.43, FOV $< 0.25$ mm) \cite{genII}.  Collected fluorescence is directed through a 45:55 beam splitter to both a Princeton Instruments Photonmax 512B EMCCD camera and a Hamamatsu H7360-02 photomultiplier tube (PMT), operating in photon counting mode.  The PMT has a quantum efficiency of .205 at $\lambda =  397$ nm.

\begin{figure}
\begin{center}
\includegraphics{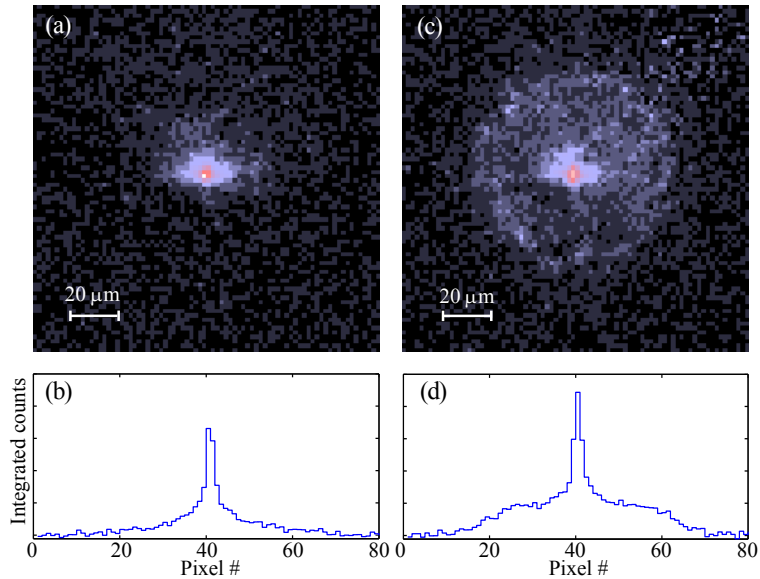}
\end{center}
\caption{\label{CCDImage} False colour CCD images of ions above (a) a trap surface, and (c) the integrated micromirror.  Background calibration images taken without an ion have been subtracted from the above images.  (b,d) Counts from the images in (a,c) integrated across each vertical line of pixels. The sharp peaks are the direct image of the ion while the wide pedestals are produced from the light reflected from the surfaces. Each pixel is 1.6 $\times$ 1.6 $\mu$m$^2$ at the magnification used in the experiment.}
\end{figure}

\begin{figure}
\begin{center}
\includegraphics{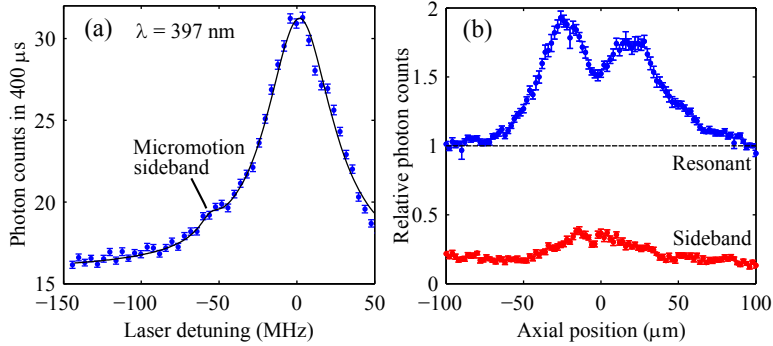}
\end{center}
\caption{\label{sideband} (a) Fluorescence versus frequency at the centre of the mirror (z=0).  Static fields have been applied to minimize the rf sideband. The remaining micromotion peak is $0.06 \pm 0.02$ of the peak intensity. This corresponds to a micromotion modulation index $\beta = 0.3\pm0.1$. The smooth curve is the least squares fit of the sum of two Lorentzian functions. (b) Relative collection intensity as a function of the ion position over the mirror. The mirror is centred roughly at 0. For each position in the scan, a field in the $\hat{x}$ direction was applied to minimize the rf sideband, thus maximizing the carrier intensity.}
\end{figure}

Figure \ref{CCDImage} compares CCD images of ions above a planar region of the trap and over the mirror.  Resonant fluorescent light directly collected from the ion appears as a well localized spot on the detector.  Photons scattered from the surface of the planar region and of the mirror face form a diffuse reflected image.  We observe a factor of 1.9 photon collection enhancement for an ion over the mirror compared to an ion above the planar region (see figure \ref{sideband}b).  To measure the dependence of the collection enhancement on ion position, we shuttle ions across the trapping zone while monitoring the fluorescence with the PMT.  A movie demonstrating ion transport across the mirror is included in the supplementary materials.  

Any rf micromotion that has a component parallel to the 397 nm cycling transition beam's propagation direction will induce rf sidebands on the fluorescence profile and reduce the on-resonant fluorescence.  Stray electric fields and small control potential errors can displace the ion from the micromotion minimum.  To counter this effect, we measure and compensate \cite{Berkeland-JApplPhys.83.5025.1998} the electric field in the radial $\hat{x}$ direction required to minimize the sideband at each point in the scan.  The corrections are less than 300 V/m and have a strong dependence on the ion position, likely due to a slight misalignment of the mirror to the trap axis. Though minimizing the sideband does not guarantee that the ion is on the micromotion minimum (there can be motion perpendicular to the laser), it does maximize the fluorescence.  Since the 397 nm beam is at an angle of 45$^\circ$ with the trap axis, the remaining micromotion seen by the laser beam is an upper bound on any axial micromotion at the ion location and the resulting pseudopotential barriers. A scan of the 397 nm laser beam frequency with the compensation applied shows only a small remnant of a micromotion sideband (see figure \ref{sideband}).

Figure \ref{sideband}b shows light collection versus ion position over the mirror with the compensating field applied.  The vertical scale gives the relative intensity collected by the PMT referenced to the intensity when the ion is sufficiently far from the mirror that the mirror no longer contributes, and a compensation is made for the slight dependence of the PMT collection efficiency on the ion position. This dependence is due to the PMT aperture and is determined by measuring the ion fluorescence versus position on a section of the trap far from the mirror. Fluorescence measurements on the micromotion sideband as a function of ion position (see figure \ref{sideband}b) show that the ion remains largely compensated at all the measurement locations even with the presence of the mirror.  The peak light collection shows a factor of 1.9 improvement as compared to collection without the mirror.  In general, the mirror reflectance has a spatial dependence from the local topography of the aluminium film. It is postulated that the drop in collected fluorescence observed directly over the mirror is related to the rough aluminium defect at the mirror centre (see section \ref{section:process} and figure \ref{cross_section}c) caused by a fabrication error. Future testing with high-quality aluminium films produced by improved fabrication techniques may clarify the role of mirror surface topography in light collection.


\section{Conclusions}
\label{section:conclusion}

We have developed a surface-electrode trap with an integrated micromirror, and observed a factor of 1.9 fluorescence collection enhancement for $^{40}$Ca$^+$ ions trapped above the mirror surface.  The trap design is optimized to improve the solid-angle coverage of the reflective optic under the constraints that no electrode edges be patterned inside the mirror cavity and that the electrode geometry remains compatible with ion shuttling.  A relay optic system has been designed to efficiently collect fluorescence over a 16 mm FOV using a multi-scale approach, enabling enhanced ion fluorescence collection over multiple mirrors distributed across a trap.

Although a significant enhancement in fluorescence collection has been demonstrated, several improvements with the current implementation may be made, including improving the profile and roughness of mirror surfaces.  Since the fabrication of the prototype described here, substantial progress has been made in the production of smoother mirror surfaces, and we expect to build mirror surfaces with $\sigma_{RMS} < 10$ nm in the near future.  With improved fabrication, we believe it is feasible to produce an integrated micromirror with greater than 85\% reflectivity at $\lambda = 397$ nm, which collects $\sim 12\%$ of emitted photons into a cone of fluorescence.  For a trap with an 85\% surface reflectivity, we estimate the collection efficiency of the system when coupled to the relay optic used in the experiment (NA $\leq$ 0.43, FOV $<$ 0.25 mm) to be 17\% over the mirror, and 9\% over a planar region ($\sim 1.8 \times$ enhancement).  The same analysis using the relay lens described in section \ref{section:relay} (NA = 0.14, FOV = 16 mm) yields estimate collection efficiencies of 12\% and 0.9\% for an ion above a micromirror and a planar region respectively ($\sim 13 \times$ enhancement).  We emphasize that a large FOV relay optic coupled to a micromirror array is expected to allow efficient, asynchronous readout over multiple ions.  We are currently considering trap designs with several integrated mirrors, and are planning experiments to demonstrate the feasibility of the multi-scale optics approach to scalable qubit detection.


\ack
\addcontentsline{toc}{section}{Acknowledgments}
We thank J Kim and T Kim of Duke University for numerous discussions regarding the integrated mirror.  JTM acknowledges the support of a Georgia Tech Presidential Fellowship.  This material is based upon work supported by the Office of the Director of National Intelligence (ODNI), Intelligence Advanced Research Projects Activity (IARPA).  All statements of fact, opinion or conclusions contained herein are those of the authors and should not be construed as representing the official views or policies of IARPA, the ODNI, or the U.S. Government.  US Army Research Office contract support through W911NF081-0315 and W911NF081-0515 is acknowledged.


\section*{References}
\addcontentsline{toc}{section}{References}



\begin{thebibliography}{10}

\bibitem{Blatt-Nature.453.1008.2008}
Blatt, R. and Wineland, D.
\newblock Entangled states of trapped atomic ions.
\newblock {\em Nature}{ \bf 453}, 1008 (2008).

\bibitem{Amini-NJP.12.033031.2010}
Amini, J.~M., Uys, H., Wesenberg, J.~H., Seidelin, S., Britton, J., Bollinger,
  J.~J., Leibfried, D., Ospelkaus, C., VanDevender, A.~P., and Wineland, D.~J.
\newblock Toward scalable ion traps for quantum information processing.
\newblock {\em New J. of Phys.}{ \bf 12}, 033031 (2010).

\bibitem{Kim-QIC.5.515.2005}
Kim, J., Pau, S., Ma, Z., McLellan, H.~R., Gates, J.~V., Kornblit, A., Slusher,
  R.~E., Jopson, R.~M., Kang, I., and Dinu, M.
\newblock System design for large-scale ion trap quantum information processor.
\newblock {\em Quantum Inf. Comput.}{ \bf 5}, 515 (2005).

\bibitem{Chiaverini-QIC.96.253003.2005}
Chiaverini, J., Blakestad, R.~B., Britton, J., Jost, J.~D., Langer, C.,
  Leibfried, D., Ozeri, R., and Wineland, D.~J.
\newblock Surface-electrode architecture for ion-trap quantum information
  processing.
\newblock {\em Quantum Inf. Comput.}{ \bf 5}, 419 (2005).

\bibitem{britton-apl.95.173102.2009}
Britton, J., Leibfried, D., Beall, J.~A., Blakestad, R.~B., Wesenberg, J.~H.,
  and Wineland, D.~J.
\newblock Scalable arrays of rf {P}aul traps in degenerate {S}i.
\newblock {\em Appl. Phys. Lett.}{ \bf 95}, 173102 (2009).

\bibitem{Seidelin-PhysRevLett.96.253003.2006}
Seidelin, S., Chiaverini, J., Reichle, R., Bollinger, J.~J., Leibfried, D.,
  Britton, J., Wesenberg, J.~H., Blakestad, R.~B., Epstein, R.~J., Hume, D.~B.,
  Itano, W.~M., Jost, J.~D., Langer, C., Ozeri, R., Shiga, N., and Wineland,
  D.~J.
\newblock Microfabricated surface-electrode ion trap for scalable quantum
  information processing.
\newblock {\em Phys. Rev. Lett.}{ \bf 96}, 253003 Jun  (2006).

\bibitem{Labaziewicz-PhysRevLett.100.013001.2008}
Labaziewicz, J., Ge, Y., Antohi, P., Leibrandt, D., Brown, K.~R., and Chuang,
  I.~L.
\newblock Suppression of heating rates in cryogenic surface-electrode ion
  traps.
\newblock {\em Phys. Rev. Lett.}{ \bf 100}, 013001 Jan  (2008).

\bibitem{Brown-PhysRevA.75.015401.2007}
Brown, K.~R., Clark, R.~J., Labaziewicz, J., Richerme, P., Leibrandt, D.~R.,
  and Chuang, I.~L.
\newblock Loading and characterization of a printed-circuit-board atomic ion
  trap.
\newblock {\em Phys. Rev. A}{ \bf 75}, 015401 Jan  (2007).

\bibitem{Leibrandt-QIC.9.0910.2009}
Leibrandt, D.~R., Labaziewicz, J., Clark, R.~J., Chuang, I.~L., Epstein, R.~J.,
  Ospelkaus, C., Wesenberg, J.~H., Bollinger, J.~J., Leibfried, D., Wineland,
  D.~J., Stick, D., Sterk, J., Monroe, C., Pai, C.-S., Low, Y., Frahm, R., and
  Slusher, R.~E.
\newblock Demonstration of a scalable, multiplexed ion trap for quantum
  information processing.
\newblock {\em Quantum Inf. Comput.}{ \bf 9}, 910 (2009).

\bibitem{Splatt-NJP.11.103008.2009}
Splatt, F., Harlander, M., Brownnutt, M., Ahringer, F.~Z., Blatt, R., and
  Ansel, W.~H.
\newblock Deterministic reordering of $^{40}${Ca}$^+$ ions in a linear
  segmented {P}aul trap.
\newblock {\em New J. Phys.}{ \bf 11}, 103008 (2009).

\bibitem{Kielpinski-Nature.417.709.2002}
Kielpinski, D., Monroe, C., and Wineland, D.~J.
\newblock Architecture of a large-scale ion trap quantum computer.
\newblock {\em Nature}{ \bf 417}, 709 (2002).

\bibitem{Burrell-PhysRevA.81.040302.2010}
Burrell, A.~H., Szwer, D.~J., Webster, S.~C., and Lucas, D.~M.
\newblock Scalable simultaneous multiqubit readout with 99.99\% single-shot
  fidelity.
\newblock {\em Phys. Rev. A}{ \bf 81}, 040302 Apr  (2010).

\bibitem{Myerson-PhysRevLett.100.200502.2008}
Myerson, A.~H., Szwer, D.~J., Webster, S.~C., Allcock, D. T.~C., Curtis, M.~J.,
  Imreh, G., Sherman, J.~A., Stacey, D.~N., Steane, A.~M., and Lucas, D.~M.
\newblock High-fidelity readout of trapped-ion qubits.
\newblock {\em Phys. Rev. Lett.}{ \bf 100}, 200502 May  (2008).

\bibitem{Barreiro-Nature.470.486.2011}
Barreiro, J.~T., M\"uller, M., Schindler, P., Nigg, D., Monz, T., Chwalla, M.,
  Hennrich, M., Roos, C.~F., Zoller, P., and Blatt, R.
\newblock An open-system quantum simulator with trapped ions.
\newblock {\em Nature}{ \bf 470}, 486 (2011).

\bibitem{Chiaverini-Nature.432.602.2004}
Chiaverini, J., Leibfried, D., Schaetz, T., Barrett, M.~D., Blakestad, R.~B.,
  Britton, J., Itano, W.~M., Host, J.~D., Knill, E., Langer, C., Ozeri, R., and
  Wineland, D.~J.
\newblock Realization of quantum error correction.
\newblock {\em Nature}{ \bf 432}, 602 (2004).

\bibitem{Raussendorf-NJP.9.199.2007}
Raussendorf, R., Harrington, J., and Goyal, K.
\newblock Topological fault-tolerance in cluster state quantum computation.
\newblock {\em New J. Phys.}{ \bf 9}, 199 (2007).

\bibitem{Maunz-PhysRevLett.102.250502.2009}
Maunz, P., Olmschenk, S., Hayes, D., Matsukevich, D.~N., Duan, L.-M., and
  Monroe, C.
\newblock Heralded quantum gate between remote quantum memories.
\newblock {\em Phys. Rev. Lett.}{ \bf 102}, 250502 Jun  (2009).

\bibitem{VanDevender-PhysRevLett.105.023001.2010}
VanDevender, A.~P., Colombe, Y., Amini, J., Leibfried, D., and Wineland, D.~J.
\newblock Efficient fiber optic detection of trapped ion fluorescence.
\newblock {\em Phys. Rev. Lett.}{ \bf 105}, 023001 Jul  (2010).

\bibitem{Streed-PhysRevLett.106.010502.2011}
Streed, E.~W., Norton, B.~G., Jechow, A., Weinhold, T.~J., and Kielpinski, D.
\newblock Imaging of trapped ions with a microfabricated optic for quantum
  information processing.
\newblock {\em Phys. Rev. Lett.}{ \bf 106}, 010502 Jan  (2011).

\bibitem{Brady-arxiv1008.2977}
Brady, G.~R., Ellis, A.~R., Moehring, D.~L., Stick, D., Highstrete, C.,
  Fortier, K.~M., Blain, M.~G., Haltli, R.~A., Cruz-Cabrera, A.~A., Briggs,
  R.~D., Wendt, J.~R., Carter, T.~R., Samora, S., and Kemme, S.~A.
\newblock Integration of fluorescence collection optics with a microfabricated
  surface electrode ion trap.
\newblock arXiv:1008.2977v2,  Sep  (2010).

\bibitem{Herskind-arxiv.1011.5259}
Herskind, P.~F., Wang, S.~X., Shi, M., Ge, Y., Cetina, M., and Chuang, I.~L.
\newblock Microfabricated surface trap for scalable ion-photon interfaces.
\newblock arXiv:1011.5259v1,  Nov  (2010).

\bibitem{Shu-PhysRevA.81.042321.2010}
Shu, G., Kurz, N., Dietrich, M.~R., and Blinov, B.~B.
\newblock Efficient fluorescence collection from trapped ions with an
  integrated spherical mirror.
\newblock {\em Phys. Rev. A}{ \bf 81}, 042321 Apr  (2010).

\bibitem{Wesenberg-PRA.78.063410.2008}
Wesenberg, J.~H.
\newblock Electrostatics of surface-electrode ion traps.
\newblock {\em Phys. Rev. A}{ \bf 78}, 063410 Dec  (2008).

\bibitem{Blakestad-PRL.102.153002.2009}
Blakestad, R.~B., Ospelkaus, C., Van{D}envender, A.~P., Amini, J.~M., Britton,
  J., Leibfried, D., and Wineland, D.~J.
\newblock High-fidelity transport of trapped-ion qubits through an {X}-junction
  trap array.
\newblock {\em Phys. Rev. Lett.}{ \bf 102}, 153002 (2009).

\bibitem{Trupke-Appl.Phys.Lett.87.211106.2005}
Trupke, M., Hinds, E.~A., Eriksson, S., Curtis, E.~A., Moktadir, Z.,
  Kukharenka, E., and Kraft, M.
\newblock Microfabricated high-finesse optical cavity with open access and
  small volume.
\newblock {\em Appl. Phys. Lett.}{ \bf 87}, 211106 (2005).

\bibitem{Albero-OpticsExpress.17.6283.2009}
Albero, J., Nieradko, L., Gorecki, C., Ottevaere, H., Gomez, V., Thienpont, H.,
  Pietarinen, J., Paivanranta, B., and Passilly, N.
\newblock Fabrication of spherical microlenses by a combination of isotropic
  wet etching of silicon and molding techniques.
\newblock {\em Opt. Express}{ \bf 17}, 6283 (2009).

\bibitem{Noek-OpticsLett.14.2460.2010}
Noek, R., Knoernschild, C., Migacz, J., Kim, T., Maunz, P., Merrill, T.,
  Hayden, H., Pai, C.-S., and Kim, J.
\newblock Multi-scale optics for enhanced light collection from a point source.
\newblock {\em Opt. Lett.}{ \bf 35}, 2460 (2010).

\bibitem{Irving-J.Electrochem.Soc.107.1020.1960}
Irving, B.~A., Robbins, H., and Schwartz, B.
\newblock Chemical etching of silicon 1. {T}he system {HF}, {HNO}$_3$, and
  {H}$_2${O}.
\newblock {\em J. Electrochem. Soc.}{ \bf 107}, 1020 (1960).

\bibitem{Svetovoy-J.Micromech.Microeng.11.2344.2007}
Svetovoy, V.~B., Berenschot, J.~W., and Elwenspoek, M.~C.
\newblock Experimental investigation of anisotropy in isotropic silicon
  etching.
\newblock {\em J. Micromech. Microeng.}{ \bf 17}, 2344 (2007).

\bibitem{IntroToFab}
Jaeger, R.~C.
\newblock {\em Introduction to Microelectronic Fabrication}.
\newblock Prentice Hall, Upper Saddle River, New Jersey,  (2002).

\bibitem{Urabe-jjap.30.1532.1991}
Urabe, S., Imajo, H., Hayasaka, K., and Watanabe, M.
\newblock Trapping of {C}a$^+$ ions and optical detection.
\newblock {\em Jpn. J. Appl. Phys.}{ \bf 30}, 1532 (1991).

\bibitem{Haffner-review}
Haffner, H., Roos, C.~F., and Blatt, R.
\newblock Quantum computing with trapped ions.
\newblock {\em Phys. Rep.}{ \bf 469}, 155 (2008).

\bibitem{genII}
Doret, S.~C., Amini, J.~M., Wright, K., Dennison, D., Faircloth, D., Killian,
  T., Volin, C., Hayden, H., Pai, C.~S., Slusher, R., and Harter, A.
\newblock Fabrication and characterization of a scalable microfabricated
  aluminum surface electrode trap.
\newblock (to be published).

\bibitem{Berkeland-JApplPhys.83.5025.1998}
Berkeland, D.~J., Miller, J.~D., Bergquist, J.~C., Itano, W.~M., and Wineland,
  D.~J.
\newblock Minimization of ion micromotion in a {P}aul trap.
\newblock {\em J. Appl. Phys.}{ \bf 83}, 5025 (1998).

\end{thebibliography}

\end{document}